\UseRawInputEncoding
\documentclass[aps,showpacs,prep,graphics]{revtex4}

\usepackage{amssymb}
\usepackage[dvips]{graphicx}
\usepackage{amsmath}
\usepackage{bm}
\usepackage{epsfig}
\usepackage{graphicx}
\usepackage{caption}
\usepackage{wrapfig}
\usepackage{color}
\usepackage{subfig}

\begin{document}

\title{Gravitational waves during Higgs inflation from complex geometrical scalar-tensor theory of gravity}

\author{ $^{1}$ Jos\'e Edgar Madriz Aguilar,  $^{2}$ A. Bernal, $^{2}$ F. Aceves de la Cruz , $^{1}$ J. A. Licea
\thanks{E-mail address: mariana.montnav@gmail.com} }
\affiliation{$^{1}$ Departamento de Matem\'aticas, Centro Universitario de Ciencias Exactas e ingenier\'{i}as (CUCEI),
Universidad de Guadalajara (UdG), Av. Revoluci\'on 1500 S.R. 44430, Guadalajara, Jalisco, M\'exico,  \\
and\\
$^{2}$ Departamento de F\'isica, Centro Universitario de Ciencias Exactas e ingenier\'{i}as (CUCEI),
Universidad de Guadalajara (UdG), Av. Revoluci\'on 1500 S.R. 44430, Guadalajara, Jalisco, M\'exico. \\
E-mail:  jose.madriz@academicos.udg.mx, alfonso.bernal@alumnos.udg.mx,
fermin.adelacruz@academicos.udg.mx,
antonio.licea@academicos.udg.mx}

\begin{abstract}

 In this paper we investigate tensor fluctuations of the metric at the end of a Higgs inflationary period in the context of a recently introduced complex geometrical scalar-tensor theory of gravity. In our model the Higgs field has a geometrical origin and the affine connection is determined by the Palatini's principle. Additionally, we consider an extra contribution to the tensor-fluctuations equation coming from the vacuum term in the energy momentum tensor associated to the Higgs field. The Higgs potential is rescaled by the non-canonicity function of the kinetic term of the field which is modified by the symmetry group of the background geometry. We obtain a nearly scale invariant spectrum and a scalar to tensor ratio in agreement with PLANCK 2018 cosmological results.
\end{abstract}

\pacs{04.50. Kd, 04.20.Jb, 02.40k, 98.80k, 98.80.Jk, 04.30w}
\maketitle

\vskip .5cm
 Weyl-Integrable geometry, Higgs scalar field, geometrical scalar-tensor gravity, inflation, gravitational waves.

\section{INTRODUCTION}

Geometrical scalar tensor theories of gravity is an approach of scalar-tensor theories of gravity that arises as an attempt to obtain an scalar invariant action under the group of symmetries of the background geometry. Specifically, when a Platini's variational principle is adopted for a scalar-tensor theory the resulting background geometry is non-riemannian and as a consequence the group of symmetries that leave invariant the non-metricity condition is bigger than the diffeomorfism group, and so the original action is not transforming as an scalar under this extended group. Thus in this geometrical approach a new action is proposed in order to be a scalar under the group of symmetries of the background geometry \cite{II1}. A previous approach in which the Palatini's principle has been incorporated in scalar-tensor theories of gravity can be found for example in \cite{CR1,CR2}. To achieve the invariance of the action under the new group of geometrical symmetries it is incorporated a gauge vector field in the covariant derivative that in certain scenarios can be identified with the electromagnetic potential. Several topics have been investigated in the framework of this approach, like for example, Higgs inflation, the formation of the seeds for cosmic magnetic fields and Dark energy scenarios \cite{II1, II2, II3}. Moreover, this geometrical approach can also be extended to a new formulation of $f(R)$ theories obtained by broken the Weyl gauge symmetry imposed by the background geometry. Cosmological backreaction consequences and CMB imprints of the new contributions have been also investigated \cite{II4}.  \\

Inflationary models can be considered as a solution to the problems of the Big Bang cosmology, whose main predictions have been verified by the acoustic peaks of CMB primordial temperature anisotropies \cite{II5}.  An important prediction of inflationary models is the relic background of gravitational waves (GW). Formally these GW are described as tensor perturbations of the metric generated by the primordial density perturbations during inflation. The fact that LIGO reported the detection of gravitational waves sourced by astrophysical objects has motivated the search of primordial GW coming from inflation. Several experiments for the detection of GW are in order, for example, the Laser Interferometer Antenna (LISA) \cite{EX1,EX2} and the DECI-hertz Interferometer Gravitational Wave  Observatory (DECIGO) \cite{EX3,EX4}. Among the zoo of inflationary models, Higgs inflationary models have attracted the interest of cosmologists because the Higgs is the unique scalar particle that has been detected. Minimal coupling Higgs inflationary models, in the context of general relativity, have the problem of reproducing the amplitude of density perturbations consistently with observational data,  which in general will depend on the quartic coupling of the Higgs potential. One way to sort the problem is the proposal of non-minimally coupled models. However, those models are not free of problems. For example, the tree-level unitarity problem that appears when radiative corrections in the standard effective Higgs potential are regarded \cite{HIN1}. The unitarity limit during inflation gives a different energy scale than the one in which the electroweak vacuum tree-level unitarity is violated, which motivates an ultraviolet extension of the model that can leave to problems with the amplitude of primordial density perturbations \cite{HIN2,HIN3}.\\

One characteristic of non-minimal coupling models of Higgs inflation is that they work in two frames: the Jordan and the Einstein frames. The pass from the Jordan to the Einstein frames is implemented by means of a conformal transformation of the metric of the form $\bar{g}_{\alpha\beta}=\Omega(h) g_{\alpha\beta}$, with $h$ being the physical Higgs field and $g_{\alpha\beta}$ the metric in the Jordan frame. The background geometry is assumed riemannian and therefore it holds that $\nabla_{\mu}g_{\alpha\beta}=0$, which is the metricity condition for this kind of geometry. However, due to the conformal transformation of the metric it is not difficult to verify that  in the Einstein frame $\nabla_{\mu}\bar{g}_{\alpha\beta}\neq 0$. So in this  frame the background geometry is no longer riemannian. This fact, has not been considered in the majority of non-minimal coupled  Higgs inflationary models, where in spite of taking the conformal transformation of the metric, the riemannian is still being considered as the background geometry in the Einstein frame.  But this is in fact considered  a very important issue in the geometrical scalar-tensor theories of gravity. The appearance of a gauge vector field that can play the role of an electromegnetic potential and the energy rescaling of the Higgs potential suggested by the symmetry group of the background geometry are examples of consequences of considering the change of the background geometry when passing from one frame to another.\\

In this paper, in the framework of complex geometrical scalar-tensor theories of gravity, we study primordial gravitational waves generated during a Higgs inflationary stage. To achieve our goal the paper is organized as follows. Section I is left for a brief introduction. Section II is devoted to the construction of the invariant action of the model in the context of geometrical scalar-tensor theories of gravity. In section III we derive the field equations of the particular Higgs inflation model. In section IV we study the tensor fluctuations of the metric in order to obtain the powwer spectrum and the scalar to tensor ratio for primordial gravitational waves at the end of inflation. Finally, section V is left for some conclusions.

\section{The action in the complex geometrical scalar-tensor theory }
Let us start with a traditional complex scalar-tensor theory, whose action can be written as \cite{II1,II2}
\begin{equation}\label{eq1}
{\cal S}=\int d^4 x\,\sqrt{-g}\, e^{-(\Phi+\Phi^{\dagger})}\left[\frac{M_p^2}{2}R+\Omega(\Phi+\Phi^{\dagger})\,\Phi^{,\mu}\Phi_{,\mu}^{\dagger}-V(\Phi+\Phi^{\dagger})\right],
\end{equation}
where $g$ denotes the determinant of the metric, $R$ is the Ricci scalar curvature, $\Omega((\Phi+\Phi^{\dagger})$ is a well-behaved differentiable function, the dagger $\dagger$ is denoting transposed complex conjugate, $V(\Phi+\Phi^{\dagger})$ is the scalar potential and $M_p=(8\pi G)^{-1/2}$ is the reduced Planckian Mass. Adopting the Palatini's variational principle the backgroun geometry associated to \eqref{eq1} is one of the Weyl-Integrable type characterized by the compatibility condition: $\nabla_{\alpha}g_{\mu\nu}=(\Phi+\Phi^{\dagger})_{,\alpha}\, g_{\mu\nu}$ with $\nabla_{\sigma}$ denoting the Weyl covariant derivative. The geometrical symmetry group that leaves invariant this condition is the Weyl group of transformations
\begin{eqnarray}\label{eq2}
\bar{g}_{\lambda\sigma}&=& e^{f+f^{\dagger}}g_{\lambda\sigma},\\
\label{eq3}
\bar{\Phi}&=& \Phi+f,
\end{eqnarray}
being $f(x^{\gamma})$ a well behaved complex function of the space-time coordinates. As the action \eqref{eq1} is not an invariant under the Weyl group, an invariant action results to be \cite{II1,II2}
\begin{equation}\label{eq4}
    {\cal S}_{inv}=\int\,d^4x\,\sqrt{-g}\,e^{-(\Phi+\Phi^{\dagger})}\left[\frac{M_p^2}{2}R+\Omega(\Phi+\Phi^{\dagger})\,\Phi^{:\mu}\Phi_{:\mu}^{\dagger}-e^{-(\Phi+\Phi^{\dagger})}V(\Phi+\Phi^{\dagger})-\frac{1}{4}e^{(\Phi+\Phi^{\dagger})}H_{\mu\nu}H^{\mu\nu}\right],
\end{equation}
where it was introduced the gauge covariant derivative $\Phi_{:\alpha}=\nabla\Phi+i\epsilon B_{\alpha}\Phi$, with $B_\alpha$ being a gauge field well-defined in the space-time manifold,  $\epsilon $ is a coupling constant and $H_{\mu\nu}=(\Phi B_{\nu})_{,\mu}-(\Phi B_{\mu})_{,\nu}$ is a field strength tensor. This is the action of a geometrical scalar-tensor theory of gravity that is different from the traditional action \eqref{eq1}. The invariance of \eqref{eq4} is achieved only when the next transformations are valid
\begin{eqnarray}\label{eq5}
\bar{\Phi}\bar{B}_{\lambda}&=&\Phi B_{\lambda}+i\epsilon^{-1}f_{,\lambda},\\
\label{eq6}
\bar{\Phi}^{\dagger}\bar{B}_{\lambda}&=& \Phi^{\dagger}B_{\lambda}-i\epsilon^{-1}f^{\dagger}_{,\lambda},\\
\label{eq7}
\bar{\Omega}(\bar{\Phi}+\bar{\Phi}^{\dagger}) &\equiv &\Omega(\bar{\Phi}+\bar{\Phi}^{\dagger}-f-f^{\dagger})=\Omega(\Phi+\Phi^{\dagger}) ,\\
\label{eq8}
\bar{V}(\bar{\Phi}+\bar{\Phi}^{\dagger}) &\equiv &V(\bar{\Phi}+\bar{\Phi}^{\dagger}-f-f^{\dagger})=V(\Phi+\Phi^{\dagger}).
\end{eqnarray}
In terms of the Weyl invariant metric: $\gamma_{\alpha\beta}=e^{-(\Phi+\Phi^{\dagger})}g_{\alpha\beta}$ and the new fields
\begin{eqnarray}\label{eq9}
    \varphi &=& \sqrt{\xi}\,e^{-\Phi},\\
    \label{eq10}
    A_{\mu} &=& B_{\mu}\ln\left(\frac{\varphi}{\sqrt{\xi}}\right),
\end{eqnarray}
the action \eqref{eq4} can be put in the form \cite{II1,II2}
\begin{equation}\label{eq11}
    {\cal S}_{inv}=\int d^4 x \,\sqrt{-\gamma}\left[\frac{M_p^2}{2}{\cal R}+\frac{1}{2}\omega(\varphi\varphi^{\dagger})D^{\mu}\varphi(D_{\mu}\varphi)^{\dagger}-\hat{V}(\varphi\varphi^{\dagger})-\frac{1}{4}F_{\mu\nu}F^{\mu\nu}\right],
\end{equation}
where $D_{\lambda}\varphi=\,\!^{(R)}\nabla_{\lambda}\varphi +i\epsilon A_{\lambda}\varphi$ is an effective Riemannian gauge covariant derivative, $F_{\alpha\beta}=A_{\beta,\alpha}-A_{\alpha,\beta}=-H_{\alpha\beta}$ is the Faraday tensor, ${\cal R}$ is the Riemannian Ricci scalar and where the next relations are valid
\begin{eqnarray}\label{eq12}
    \Phi_{:\sigma}&=&-\frac{1}{\varphi}D_{\sigma}\varphi,\\
    \label{eq13}
    \omega(\varphi\varphi^{\dagger})&=&\frac{2\,\Omega\left[\ln(\varphi\varphi^{\dagger}/\xi)\right]}{\varphi\varphi^{\dagger}},\\
    \label{eq14}
    \hat{V}(\varphi\varphi^{\dagger}) &=& V \left(\ln\frac{\varphi\varphi^{\dagger}}{\xi}\right),
\end{eqnarray}
with $\xi$ being a constant parameter introduced in order  to the field $\varphi$ has the correct physical units. 

\section{The field equations of the Higgs inflationary model}

In order to propose a Higgs inflationary model we start considering the Higgs potential
\begin{equation}\label{eq15} V(\Phi\Phi^{\dagger})=\frac{\lambda}{4}\left(\Phi\Phi^{\dagger}-\sigma^2\right)^2,
\end{equation}
with $\lambda=0.129$ and the vacuum expectation value $\sigma=246\,GeV$ \cite{III1,III2}. In terms of the field $\varphi$ the expression \eqref{eq15} reads
\begin{equation}\label{eq16}
V(\varphi\varphi^{\dagger})=\frac{\lambda}{4}\left(\frac{\varphi\varphi^{\dagger}}{\xi}-\sigma^2\right)^2.
\end{equation}
On the other hand, the action \eqref{eq11} has a Riemannian background geometry and thus the fields $\varphi$ and $A_{\mu}$ respect the gauge transformations
\begin{eqnarray}\label{eq17}
    \bar{\varphi}=\varphi\,e^{i\epsilon \theta(x)},\\
    \label{eq18}
    \bar{A}_{\nu}=A_{\nu}-\theta_{,\mu},
\end{eqnarray}
where $\theta (x)$ is a well-behaved gauge function. In this manner, breaking the symetry by taking $\varphi=\varphi^{\dagger}$ and with the gauge election $\theta_{,\nu}=A_{\nu}$, the action \eqref{eq11} acquires the form
\begin{equation}\label{eq19}
    {\cal S}=\int d^{4}x \sqrt{-\gamma}\,\left[\frac{M_p^2}{2}{\cal R}+\frac{1}{2}\omega_{eff}({\cal H}){\cal H}^{,\mu}{\cal H}_{,\mu}-V_{eff}({\cal H})\right],
\end{equation}
where ${\cal H}$ is the Higgs field that obeys: $\varphi(x^{\alpha})=\sqrt{\xi}\,\sigma +{\cal H}(x^{\alpha})$, and $V_{eff}({\cal H})=V[\sqrt{\xi}\,\sigma +{\cal H}(x^{\alpha})]$. Unitarizing the kinetic term in \eqref{eq19} we arrive to
\begin{equation}\label{eq20}
  {\cal S}=\int d^{4}x \sqrt{-\gamma}\,\left[\frac{M_p^2}{2}{\cal R}+\frac{1}{2}\phi^{,\mu}\phi_{,\mu}-U(\phi)\right],  
\end{equation}
where 
\begin{eqnarray}\label{eq21}
    \phi(x^{\alpha})&=&\int \sqrt{\omega_{eff}({\cal H})}\,d{\cal H},\\
    \label{eq22}
    U(\phi)&=&V_{eff}({\cal H}(\phi))=\frac{\lambda}{4}\left[\frac{(\sqrt{\xi}\,\sigma+{\cal H}(\phi))^2}{\xi}-\sigma^2\right]^{2}.
\end{eqnarray}
The field equations resulting from \eqref{eq20} are then
\begin{eqnarray}\label{eq23}
    {\cal R}_{\mu\nu}-\frac{1}{2}{\cal R}\,\gamma_{\mu\nu}&=&M_p^{-2}T_{\mu\nu},\\
    \label{eq24}
    \Box\phi+U^{\prime}(\phi)&=&0,
\end{eqnarray}
where the energy-momentum tensor for the scalar field $\phi$ is given by
\begin{equation}\label{eq25}
    T_{\mu\nu}=\phi_{,\mu}\phi_{,\nu}-\frac{1}{2}\gamma_{\mu\nu}\left(\phi^{,\alpha}\phi_{,\alpha}-2U(\phi)\right),
\end{equation}
and $\Box$ is denoting the D'Alambertian operator. \\

Now, in order to allow the potential \eqref{eq22} to show a plateu for  large enough  field values, suitable to describe a period of inflation, we consider the anzats
\begin{equation}\label{eq26}
    \omega_{eff}({\cal H})=\frac{1}{\left[1-\beta^2(\sqrt{\xi}\,\sigma +{\cal H}^4)\right]^{5/2}},
\end{equation}
with $\beta$ being a constant parameter with $M_p^{-2}$ units. By using \eqref{eq21}, the equation \eqref{eq26} implies a relation between the inflaton field $\phi$ and the Higgs field ${\cal H}$ of the form
\begin{equation}\label{eq27}
    \phi=\frac{\sqrt{\xi}\,\sigma+{\cal H}}{[1-\beta^2(\sqrt{\xi}\sigma+{\cal H})^4]^{1/4}}.
\end{equation}
Thus, the inflationary potential \eqref{eq22} reads
\begin{equation}\label{eq28}
    U(\phi)=\frac{\lambda}{4\xi^2}\left(\frac{\phi^4}{1+\beta^2\phi^4}\right).
\end{equation}
A similar potential is obtianed for example in  \cite{III3}. Once inflation  starts 
 it is verified that $\beta^2\phi^4\ll 1$ and hence the potential can be approximated by $U(\phi)\simeq (\lambda/4\xi^2)\phi^4$.

\section{Tensor fluctuations of the metric}

The background of gravitational waves generated at the end of inflation are due to sourceless tensor fluctuations of the metric. However, the energy-momentum tensor \eqref{eq25} can be formally decomposed as a pressureless material and a vacuum parts \cite{IV1}. The vacuum component is given by
\begin{equation}\label{eq29}
    T_{\mu\nu}^{(vac)}=\left(U(\phi)-\frac{1}{2}\phi^{,\alpha}\phi_{,\alpha}\right)\gamma_{\mu\nu}.
\end{equation}
The perturbed line element has the form
\begin{equation}\label{eq30}
    ds^2=dt^2-a^2(t)\left(\delta_{ij}+h_{ij}\right)dx^{i}dx^{j}\quad ,
\end{equation}
where the tensor fluctuations of the metric are describe by $h_{ij}(x^{\alpha})$, that satisfies $tr(h_{ij})=0$ and $h_{ij}^{\,,i}=0$.
Hence, it follows from \eqref{eq23} that tensor fluctuations of the metric obey the dynamical equation
\begin{equation}\label{eq31}
    \delta {\cal R}_{\mu\nu}-\frac{1}{2}{\cal R}^{(b)}\delta\gamma_{\mu\nu}-\frac{1}{2}\delta {\cal R}\,\gamma_{\mu\nu}^{(b)}=M_{p}^{-2}\,T_{\mu\nu}^{(vac)}(\phi_b),
\end{equation}
where we have employed a semiclassical aproximation for the inflaton field that reads
\begin{equation}\label{eq32}
    \phi(x^{\lambda})=\phi_{b}(t)+\delta\phi(x^{\lambda}),
\end{equation}
where the espectation values  $<\phi>=\phi_b$ and $<\dot{\delta\phi}>=0$, being $\phi_{b}(t)$ the background inflaton field defined on cosmological scales and $\delta\phi$ describing the quantum fluctuations of the inflaton on small scales. Thus $\gamma_{\mu\nu}^{(b)}$ is the background metric, ${\cal R}^{(b)}$ accounts for the Ricci scalar evaluated on the background metric and $\delta{\cal R}$ represents the fluctuations of the Ricci scalar generated by the perturbed metric in \eqref{eq30}. \\

Thus, with the use of \eqref{eq30} the energy density for the vacuum scalar field coming from \eqref{eq29} results
\begin{equation}\label{eq33}
    \rho_{vac}=-\left[\frac{1}{2}\dot{\phi}_b^2-U(\phi_b)\right].
\end{equation}
In order to obtain a positive $\rho_{vac}$ necessarily $U(\phi_b)>(1/2)\dot{\phi}_b^2$, so the slow-roll condition on the inflaton field must be valid. The Ricci scalar has no contributions of the first order of tensor metric fluctuations and its background value is given by
\begin{equation}\label{eq34}
    {\cal R}^{(b)}=-6\left(\dot{H}+2H^2\right),
\end{equation}
with $H(t)$ being the Hubble parameter.
Thus, in the traceless-transverse (TT) gauge and in the slow-roll regime, it follows from \eqref{eq31} that the dynamics of the tensor modes is given by the linearized equations
\begin{equation}\label{eq35}
    \delta {\cal R}_{ij}-\frac{1}{2}{\cal R}^{(b)}\,\delta\gamma_{ij}=M_{p}^{-2}\,U(\phi_b)\,\delta\gamma_{ij}.
\end{equation}
With the help of \eqref{eq30} the expression \eqref{eq35} reduces to
\begin{equation}\label{eq36}
  \ddot{h}^{i}_{j}+3H\dot{h}^{i}_{j}-\frac{1}{a^2}\nabla^2h^{i}_{j}-2(2\dot{H}+3H^2)h^{i}_{j}+\frac{2}{M_p^2}U(\phi_b)h^{i}_{j}=0.  \end{equation}
  On the other hand, it follows from \eqref{eq23} and \eqref{eq24} that the background dynamics is given by
  \begin{eqnarray}\label{eq37}
      && 3H^2=M_{p}^{-2}\,U(\phi_b),\\
      \label{eq38}
&& \ddot{\phi}_b+3H\dot{\phi}_b+U^{\prime}(\phi_b)=0.
  \end{eqnarray}
By using \eqref{eq28} for $\beta^2\phi_{b}^4\ll 1$ in \eqref{eq37} we obtain a scale factor of the form \cite{II1}
\begin{equation}\label{39}
    a(t)=a_e\exp\left[\frac{\phi_e^2}{8M_p^2}\left(1-\exp\left(4M_p\sqrt{\frac{\lambda}{3\xi^2}}(t_e-t)\right)\right)\right],
\end{equation}
which at the end of inflation becomes
\begin{equation}\label{eq40}
    a(t)\simeq a_e\exp \left(-\frac{\phi_e^2}{2M_p}\sqrt{\frac{\lambda}{3\xi^2}}t_e\right)\exp\left(\frac{\phi_e^2}{2M_p}\sqrt{\frac{\lambda}{3\xi^2}}t\right),
\end{equation}
where $t_e$ denotes the time at the end of inflation,  $a_e=a(t_e)$ and $\phi_e=\phi_b(t_e)$. 
Following the canonical quantization procedure we implement the Fourier expansion
\begin{equation}\label{eq41}
    h^{i}_{j}(t,\bar{r})=\frac{e^{-\frac{3}{2}\int H(t)dt}}{(2\pi)^{3/2}}\int d^3k\,\sum_{\alpha=+,\times}\,\!^{(\alpha)}e^{i}_{j}\left[a_k^{(\alpha)}e^{i\bar{k}\cdot\bar{r}}\xi_k(t)+a_{k}^{(\alpha)\,\dagger}e^{-i\bar{k}\cdot\bar{r}}\xi_k^{*}(t)\right]
\end{equation}
with the creation $a_k^{(\alpha)\,\dagger}$ and annihilation $a_{k}^{(\alpha)}$ operators obeying the algebra
\begin{eqnarray}
    \label{eq42} 
    \left[a_k^{(\alpha)},a_{k^{\prime}}^{(\alpha^{\prime})\,\dagger}\right] = \gamma^{\alpha\alpha^{\prime}}\delta^{(3)}\left(\bar{k}-\bar{k}^{\prime}\right), \\
    \label{eq43}
    \left[a_k^{(\alpha)},a_{k^{\prime}}^{(\alpha^{\prime})}\right]=\left[a_{k}^{(\alpha)\,\dagger},a_{k^{\prime}}^{(\alpha^{\prime})\,\dagger}\right]=0,
    \end{eqnarray}
    and where the polarization tensor $e_{ij}$ satisfies the properties
    \begin{eqnarray} \label{eq44}
&& ^{(\alpha)e_{ij}}=\,\!^{(\alpha)}e_{ji},\quad k^{i}\,\!^{(\alpha)}e_{ij}=0,\\
      \label{eq45}
      && ^{(\alpha)}e_{ii}=0,\quad ^{(\alpha)}e_{ij}(-\bar{k})=\,\!^{(\alpha)}e_{ij}^{*}(\bar{k}).
       \end{eqnarray}
Now, following the canonical quantization procedure we impose the commutation relation
\begin{equation}\label{eq46}
  \left[h^{i}_{j}(t,\bar{r}),\Pi_{i}^{i}(t,\bar{r}^{\prime})\right]  =i\delta^{(3)}\left(\bar{r}-\bar{r}^{\prime}\right),
\end{equation}
where $\Pi_{ij}=\partial L/\partial \dot{h}^{ij}$ is the canonical conjugate momentum. The lagrangian for gravitational tensor modes has the form
\begin{equation}\label{eq47}
    L=\frac{M_p^2a^3}{8}\left[\dot{h}_{ij}^2-\frac{1}{a^2}h_{ij,l}h^{ij,l}+\left(2(2\dot{H}+3H^2)-\frac{2}{M_p^2}\right)h_{ij}h^{ij}\right].
\end{equation}
Thus \eqref{eq47} reduces to
\begin{equation}\label{eq48}
   \left[ h_{j}^{i}(t,\bar{r}),\dot{h}^{i}_{j}(t,\bar{r}^{\prime})\right]=\frac{4i}{a^3M_p^2}\delta^{(3)}(\bar{r}-\bar{r}^{\prime}).
\end{equation}
Now, inserting \eqref{eq41} in \eqref{eq48} we obtain
\begin{equation}\label{eq49}
    \xi_k\dot{\xi}^{*}-\xi_k^{*}\dot{\xi}_k=\frac{4i}{M_p^2a_o^3},
\end{equation}
which is the normalization condition for the modes. With the help of \eqref{eq36} and \eqref{eq41} the modes at the end of inflation are governed by the dynamical equation
\begin{equation}\label{eq50}
    \Ddot{\xi_k} + \left[\frac{{k^2}}{a_e^2e^{-2H_et_e}} e^{-2H_e t}-  \frac{33}{4} H_e^2 + \frac{2}{M_p^2} U_e\right] \xi_k = 0,
\end{equation}
where 
\begin{equation}\label{eq51}
     U_e = \frac{\lambda}{4 \xi^2} \left( \frac{\phi_e^4}{1+\beta^2 \phi_e^2}\right),
\end{equation}
and where we have employed \eqref{eq40}, with 
\begin{equation}\label{eq52}
    H_e = \frac{\phi_e^2}{2M_p}\sqrt{\frac{\lambda}{3 \xi^2}}\,.
\end{equation}
By means of \eqref{eq49} and considering the Bunch-Davies vacuum, the normalized solution of \eqref{eq50} is given by
\begin{equation}\label{eq53}
    \xi_k (t) =\frac{1}{M_p}\sqrt{\frac{\pi}{\Tilde{a_e}^3 H_e}} \mathcal{H}_{\nu}^{(2)}\left[Z(t)\right],
\end{equation}
where ${\mathcal H}_{\nu}^{(2)}[Z(t)]$ denotes the second kind Hankel function and 
\begin{eqnarray}\label{eq54}
  \nu &=& \frac{1}{H_e}\sqrt{\frac{33}{4}H_e^2-\frac{2U_e}{M_p^2}},\\  
  \label{eq55}
  Z(t) &=& \frac{k}{\Tilde{a_e} H_e} e^{-H_e t},
\end{eqnarray}
with $\Tilde{a}_e=a_e\exp\left(-H_et_e\right)$. \\

In this manner, the amplitude of gravitational waves  defined by $<h^2>_{IR}=<0|h^{i}_{j}h_{i}^{j}|0>$, on the IR-sector, i.e. on cosmological scales, is given by
\begin{equation}\label{eq56}
    \langle h^2 \rangle_{IR} = \frac{e^{- \int 3 H dt}}{2\pi^2} \int_{0}^{\epsilon k_H} \frac{dk}{k} k^3 \left(\xi_k \xi_k^*\right)\Bigr|_{IR},
\end{equation}
where $\epsilon = k_{max}^{IR} / K_p \ll 1$ is a dimensionless parameter, being  $k_{max}^{IR} = k_H(t_r)$ the wave number associated with the Hubble radius at the time $t_r$ when the modes re-enter to the horizon near the end of inflation, $k_p$ is the Planckian wave number. For example, $\epsilon$ varies from $10^{-5}$ to $10^{-8}$ for a typical Hubble parameter during inflation of the order $H\simeq 0.5\times 10^{-9}\,M_p$, which corresponds to a number of e-foldings $N\simeq 63$.\\

\begin{figure}[h]
    \centering
    \subfloat[This is a plot of the variation of $n_s$ with respect $U_e$.]{
    \label{fig1}
    \includegraphics[width=0.4\textwidth]{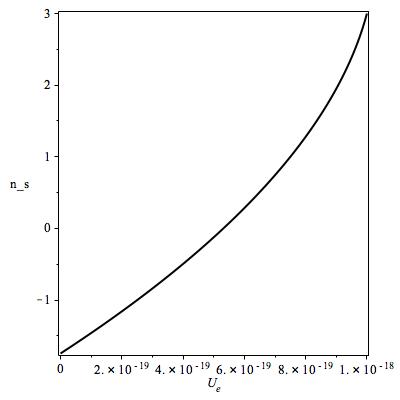}}
    \subfloat[This is a plot exhibiting the behavior of $r$ versus $\phi_b$ for different values of $\beta$.]{
    \label{fig2}
    \includegraphics[width=0.49\textwidth]{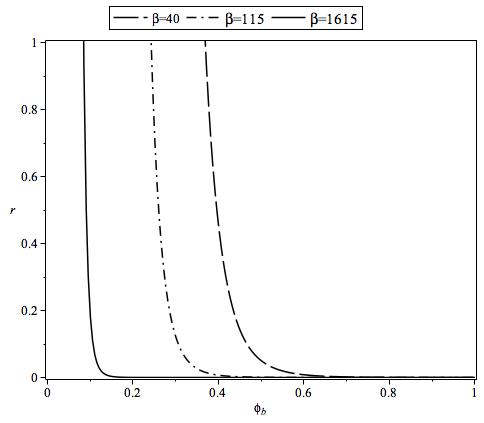}}
     \caption{These plots shows the behaviour of the spectral index $n_s$ and the scalar to tensor ratio $r$, respectively. In figure (a) the observational value for $n_s$ is achieved for $7.4258\times 10^{-19}M_p^4<U_e< 7.4417\times ^{-19} M_p^4$. 
     In figure (b) when $\beta$ increases the value of  $\phi_b$ for which $r$ enters in the observational range $r<0.056$,  decreases.}
    \label{fig:figure}
\end{figure}

Now, on cosmological scales and at the end of the inflationary period we can employ the IR-asymptotic aproximation formula 
\begin{equation}
    \label{eq57}
    {\cal H}^{(1)}_{\nu}[Z]\simeq \frac{i}{\pi}\Gamma(\nu)\left(\frac{Z}{2}\right)^{-\nu}.
\end{equation}
Hence it follows from \eqref{eq53} to \eqref{eq56} that
\begin{equation}\label{eq58}
    \langle h^2 \rangle_{IR} = \frac{2^{2\nu}}{ \pi^3} \frac{\Gamma^2 (\nu)}{M_p^2} \frac{H_e^2}{(\Tilde{a_e}H_e)^{3-2\nu}}e^{(2\nu-3)H_e  t} \int_{0}^{\epsilon k_H} \frac{dk}{k} k^{3-2\nu},
\end{equation}
where according to the modes equation \eqref{eq50} the wave number associated to the horizon is given by
\begin{equation}\label{eq59}
    k_H = \Tilde{a_e}\sqrt{\frac{11}{2}\Dot{H} + \frac{33}{4} H_e^2 + \frac{2}{M_p^2} U_e}.   
\end{equation}
In this manner we obtain a power spectrum derived from \eqref{eq58} of the form
\begin{equation}\label{eq60}
    P_h (k) = \frac{2^{2\nu+2}}{ \pi} \frac{\Gamma^2 (\nu)}{M_p^2} \left(\frac{H_e}{2\pi}\right)^2 e^{(2\nu-3)H_e  t}  \left(\frac{k}{\Tilde{a_e}H_e}\right)^{3-2\nu}.
\end{equation}
It is not difficult to verify that a nearly scale invariant power spectrum of the Harrison Zeldovick type can be achieved from \eqref{eq54} and \eqref{eq60} when $U_e\simeq 3M_p^2H_e^2$. In this particular case the formula \eqref{eq60} reduces to
\begin{equation}
    \label{eq61}
    P_h(k) |_{\nu \simeq 3/2}\simeq \frac{2^3}{M_p^2} \left( \frac{H_e}{2\pi}\right)^2.
\end{equation}
The spectral index is given by
\begin{equation}\label{eq62}
    n_s=4-\frac{2}{H_e}\sqrt{\frac{33}{4}H_e^2-\frac{2U_e}{M_p^2}}.
\end{equation}
The Planck 2018 results indicate that for the spectral index the limits are: $n_s=0.9649\pm 0.0042$ \cite{PRIC}. In the figure [\ref{fig1}] we show the behavior of the spectral index $n_s$ versus $U_e$. It can be seen that the observational values for $n_s$ can be obtained for $U_e\in\left[7.4258\times 10^{-19}, 7.4417\times ^{-19}\right]M_p^4$. \\

On the other hand, as it was shown in \cite{II1}, the power spectrum associated to the quantum fluctuations of the inflaton is given by 
\begin{equation}\label{eq63}
   P_{\delta \phi}(k) = \frac{2^{2\nu-1}}{ \pi} \Gamma^2(\nu) \left( \frac{H_e}{2\pi}\right)^2 e^{(2\nu-3) H_e t_e} \left( \frac{k}{\Tilde{a_e} H_e}\right)^{3-2\nu}.  
\end{equation}
Thus the power spectrum for curvature perturbations $P_\mathcal{R} (k)  = \frac{1}{2 \epsilon} \frac{ P_{\delta \phi}}{M_p^2}$ results
\begin{equation}\label{eq63}
    P_\mathcal{R} (k)  = \frac{2^{2\nu}}{ 4 \pi \epsilon} \frac{\Gamma^2 (\nu)}{M_p^2} \left(\frac{He}{2\pi}\right)^2 e^{(2\nu-3)H_e  t}  \left(\frac{k}{\Tilde{a_e}H_e}\right)^{3-2\nu},
\end{equation}
where we have employed de slow-roll parameter $\epsilon = \frac{M_p^2}{2} \left(\frac{U'}{U}\right)^2$. Hence the scalar to tensor ratio $ r = \frac{P_h}{P_\mathcal{R}}$ in terms of the background inflaton field has the form
\begin{equation}\label{eq65}
    r = \frac{128 M_p^2}{\phi_b^2(1 + \beta^2 \phi_b^4)^2}.
\end{equation}
According to the Planck 2018 results $r<0.056$ \cite{PRIC}. In the figure [\ref{fig2}] we show a plot of $r$ vs $\phi_b$ for different values for $\beta$. In general depending of the $\beta$ parameter the observational range of values for $r$ is achieved for an interval of values for $\phi_b$. For example, when $\beta=40\,M_p^{-2}$ the scalar to tensor ratio results $r=0.04$ when $\phi_b=0.511\,M_p$. Thus for this value of $\beta$ the observational range $r<0.056$ is achived when $\phi_b > 0.4945\,M_p$. Hence, as it is shown in figure [\ref{fig2}] for increasing values of $\beta$ the observational range for $r$ is reached for decreasing values of $\phi_b$.

\section{Conclusions}

In this paper we have investigated the background relic of gravitational waves generated during a Higgs inflationary stage  in the context of a geometrical scalar-tensor theory of gravity. In this model the Higgs scalar field has a geometrical origin because the background geometry is of the Weyl-integrable type where the Weyl scalar field is related to the Higgs field. The background geometry is asigned by the Palatini's variational principle. The physical field equations for the inflaton Higgs scalar field are obtained by  recasting the original action of the theory in terms of the so called invariant action.  \\

The primordial gravitational waves are described by tensor fluctuations of the metric. In general these kind of fluctuations are considered sourceless. One important difference with respect other approaches is that in our model we have decomposed the energy momentum-tensor into a pressureless matter plus a vacuum components, and hence the vacuum part has been considered in the formulation of the dynamical equation that governs the tensor fluctuations of the metric. \\

As a consequence of taking into account the symmetry group of the Weyl-Integrable background geometry of the original action, the standard Higgs potential is rescaled by means of a function that makes the kinetic term non-canonical, which is determined by two factors: the Weyl group of symmetries and the requirement that such function must create an enough plateu in the effective potential to achieve an initial Hubble parameter of the order $H_0\simeq 10^{11}-10^{12}$ GeV, which allows to have the enough quantity of inflation in agreement with PLANCK data \cite{PRIC,UL1}. Aditionally, the slow-roll conditions that are typically imposed, here are obtained through the requirement  that the energy density of the background inflaton field to be positive.  Hence, the rescaled potential ends up depending on the parameter $\beta$ (see eq.\eqref{eq28}). We obtain a power spectrum for gravitational waves nearly scale invariant in agreement with PLANCK data for a range of values for the effective potential at the end of inflation $U_e$ given for $[7.4528\times 10^{-19},7.4417\times 10^{-19}]M_p^4$.
The amplitude of the  spectrum results proportional to $(H/2\pi)^2$ with $H$ evaluated at the end of inflation. The scalar to tensor ratio $r$ resulted to be depending on the $\beta$ parameter. Thus for increasing values of $\beta$ the observational values of $r$ ($r<0.056$ according to PLANCK data ) are reached for decreasing values of $\phi_b$ (see figure [\ref{fig2}]).
For example, for $\beta=40\,M_p^{-2}$ we obtain $r=0.04$.

\section*{Acknowledgements}

\noindent J. E. Madriz-Aguilar, A. Bernal, F. Aceves and J. A. Licea acknowledge  CONACYT
M\'exico and Centro Universitario de Ciencias Exactas e Ingenierias of Guadalajara University for financial support.




\end{document}